\documentclass[aps,prl,twocolumn,showpacs,superscriptaddress]{revtex4}
\usepackage{graphicx}
\usepackage{dcolumn}
\usepackage{bm}
\usepackage{epstopdf}

\begin{document}

\title{Investigations of local electronic transport in InAs nanowires by 
scanning gate microscopy at helium temperatures.}

\author{A.A.~Zhukov}
\affiliation{Institute of Solid State Physics, Russian Academy of
Science, Chernogolovka, 142432 Russia}

\author {Ch.~Volk}
\author {A.~Winden}
\author {H.~Hardtdegen}
\affiliation{Peter Gr\"unberg Institut (PGI-9), Forschungszentrum
J\"ulich, 52425 J\"ulich, Germany} \affiliation{ JARA-Fundamentals of
Future Information Technology, Forschungszentrum J\"ulich, 52425
J\"ulich, Germany}
\author {Th.~Sch\"apers}
\affiliation{Peter Gr\"unberg Institut (PGI-9), Forschungszentrum
J\"ulich, 52425 J\"ulich, Germany} \affiliation{ JARA-Fundamentals of
Future Information Technology, Forschungszentrum J\"ulich, 52425
J\"ulich, Germany} \affiliation{II. Physikalisches Institut, RWTH
Aachen University, 52056 Aachen, Germany}
\date{\today}

\begin{abstract}
In the current paper a set of experiments dedicated to investigations of local electronic transport in undoped InAs nanowires at helium temperatures in the presence of a charged atomic-force microscope tip is presented. Both nanowires without defects and with internal tunneling barriers were studied. The measurements were performed at various carrier concentrations in the systems and opacity of contact-to-wire interfaces. The regime of Coulomb blockade is investigated in detail including negative differential conductivity of the whole system. The situation with open contacts with one tunneling barrier and undivided wire is also addressed. Special attention is devoted to recently observed quasi-periodic standing waves.
\end{abstract} 
\pacs{73.23.Hk, 73.40.Gk, 73.63.Nm}

\maketitle

\section{1. Introduction.}

In the past decade considerable interest was devoted to investigations of electronic transport of individual one-dimensional mesoscopic systems such as carbon nanotubes (CNT) and nanowires (InAs, InN) \cite{ref16, ref15, ref17, refBloe, refWirt, ref22, zhukov1, zhukov2}. Usually, for such experiments field-effect transistors are fabricated from individual one-dimensional (1D) objects by forming metallic contacts by electron-beam lithography. The highly doped Si substrate is used as a gate to alter the main parameters such as the carrier density and the opacity of the tunneling barriers in the metallic contacts to 1D object interfaces. The issue of contact opacity at the interfaces is extremely important. In fact, it may change the whole scenario from well defined Coulomb blockade \cite{Makaro} in case of closed contact tunneling barriers to Fabry-Perot interference for almost opened contacts \cite{FPCNT}.

Generally speaking, one-dimensional systems are rather different and may have as ballistic (CNT) \cite{FPCNT} so diffusive dominating transport as in InAs \cite{ref16, ref17, refBloe, refWirt} or InN nanowires \cite{ref22}. Additionally, crystal structure defects may result in the formation of tunneling barriers dividing the tube or wire into sequentially connected quantum dots and by that suppress the conductivity of the whole system dramatically \cite{Bleszynski2005}. Thus, the ability of altering of the potential profile locally and influence the density of carriers in 1D system locally is rather desirable. 

One viable ways to realize such local impact on a 1D system is to create a set of side- or top-gates along the nanowire or nanotube \cite{Fasth}. A more versatile and elegant method which became popular in the past decades is to use a charged tip of an atomic-force microscope as a mobile gate, so-called scanning gate microscopy (SGM). Using SGM different types of low-dimensional micro- and nano-structures have been investigated: quantum point contacts  \cite{TopinkaS2000, TopinkaN2001, refSchnezQPC2011}, quantum rings \cite{Hackens2006}, and quantum dots based on heterojunctions \cite{Pioda2004, Gildemeister2007, Huefner2011}, graphene \cite{refSchnez2010}, and carbon nanotubes \cite{Bockrat2001, Woodside2002, Zhukov2009}. The additional benefit of this technique is the resulting clear and easy reading and intuitively understandable images of quite high resolution (up to ~100 nm), thus shedding light on internal local structure of system under investigation.

In the current manuscript a set of SGM experiments, performed on InAs nanowires at $T=$4.2K, is presented. Experiments are carried out at different carrier densities, different opacities of the contacts and on wires with different numbers of internal tunneling barriers. Thus, a wide range of scenarios from closed sequentially connected quantum dots demonstrating Coulomb blockade to metallic diffusive wire are realized and investigated. Special attention is devoted to induced negative differential conductance (NDC) \cite{Zhukov2012} and recently observed quasi-periodic oscillations \cite{Zhukov2012SW, Zhukov2014SWComp}.  

This article is organized as follows: in section 2 experimental details are described, in section 3.1 experimental results obtained on InAs nanowires in Coulomb blockade regime are presented and discussed, section 3.2 is dedicated to experiments investigating recently observed quasi-periodic oscillations and the conclusion is made in section 4.

\section{2. Experimental details.}

In all our experiments we study nominally undoped InAs nanowires grown by selective-area metal-organic vapor-phase epitaxy \cite{Akabori2009}. The diameter of the wires is 100\,nm, typically. The wire is placed on an $n$-type doped Si (100) substrate covered by a 100~nm thick SiO$_2$ insulating layer. The Si substrate serves as a back-gate electrode. The evaporated Ti/Au contacts to the wire as well as the markers of the search pattern were defined by electron-beam lithography. The distance $l_{wire}$ between the contacts varies for different samples from $1.5\,\mu$m to $3.5\,\mu$m. A scanning electron beam micrograph of the typical sample is shown in Fig.~1a). The source and drain metallic electrodes connected to the wire are marked by S and D.
\begin{figure}
\includegraphics*[width=0.95\columnwidth]{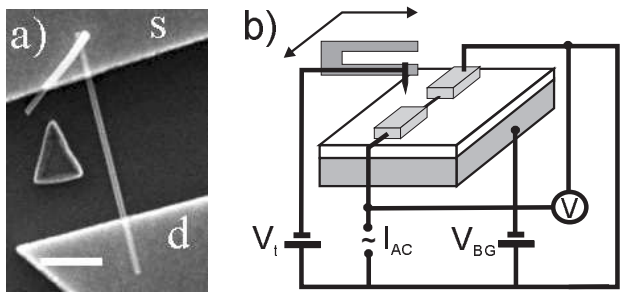}
\caption{a) Scanning electron micrograph of the typical InAs wire with metallic contacts. The source and drain contact pads are marked by 's' and 'd'. The horizontal scale bar corresponds to 1~$\mu$m. The metallic triangle on the left side of the wire is a marker of the search pattern.
b) Electrical circuit of the scanning gate microscopy measurements.
The back-gate voltage $V_{BG}$ is applied to the doped Si substrate while the tip voltage $V_t$ is applied to the tungsten tip of the probe microscope.
The resistance of the wire is measured using a two terminal circuit by applying a current ($I_{AC}$) and measuring the voltage ($V$). From Ref. \cite{Zhukov2012SW}.}
\label{Fig1}
\end{figure}

Investigations of grown InAs wires by transmission electron microscopy demonstrate that the wires are usually covered by a 2-nm-thick oxide layer  \cite{ref16}. In order to create good ohmic contacts to wire, this oxide must be removed. But even with such precautions some residual Schottky-like potential barriers may be observed even at room temperatures \cite{ref37}. These barriers becomes essentially less visible at high back gate voltages and this behavior is quite similar to one on the interface of CNT and Pd metallic contacts \cite{Makaro}.

All measurements presented in this paper are performed at $T=\,4.2$\,K. In our experimental setup we have several parameters which can been changed independently such as the tip voltage ($V_{t}$), the back gate voltage ($V_{BG}$) and the position of the tip, see Fig. 1b). Additional parameters dedicated to measurements of the conductance of the wire in a two-terminal circuit are source-to-drain DC voltage and driving AC current/voltage. The measurements are done by using a standard lock-in technique. If the wire demonstrates strong Coulomb blockade regime an AC driving voltage is applied and the resulting current is measured. In the opposite case of open contacts a driving AC current is applied, while the voltage is measured by a differential amplifier. The detailed values of the driving AC voltage and current are adjusted for a certain experimental setup and sample, and can be found elsewhere \cite{Zhukov2011P,Zhukov2012,Zhukov2012SW,Zhukov2013,Zhukov2014SWComp}.

In the SGM experiments the charged tip of a home-built scanning probe microscope \cite{AFM} is used as a mobile gate during scanning gate imaging measurements keeping the potential of the scanning probe microscope tip as well as the back-gate voltage and driving AC current/voltage constant. Another type of experiments presented here (see section 3) is based on the frozen tip position and tip voltage while $V_{BG}$ and DC source-to-drain voltages are swept for mapping of the conductance $V_{SD}$ vs. $V_{BG}$ typical for measurement in the Coulomb blockade regime.

An additional parameter in SGM experiments is the tip to SiO$_2$ surface distance. Its crucial effect on the sample conductance and thus on resulting experimental SGM images will be discussed in detail in section 3.2. 

\section{3. Experimental results and discussion.}

\subsection{3.1. Electronic transport in nanowires with defects: Coulomb blockade regime.}

In this section a set of experimental results obtained in InAs nanowires containing one or more tunneling barriers is presented and discussed. Generally speaking, the potential profile of InAs wire along the wire axis is not flat even if the wire lies on the thin (100~nm) SiO$_2$ layer and the screening from the back-gate is quite comprehensive. Essential crystal structure defects result in the tunneling barriers of different opacities. The opacity of the barriers increases with increasing the back-gate voltage. This happens partially because the decreasing widths of depletion regions arranging the barrier from both sides. Some of barriers may produce nonlinear transport even at quite high back-gate voltages ($V_{BG}=7.5$~V) (Fig.~3b in \cite{Zhukov2011P}).

If the opacity of barrier is low the wire becomes divided into sequentially connected quantum dots demonstrating Coulomb blockade at $T=4.2$~K. Detailed investigations of electronic transport of such InAs nanowires in linear and non-linear transport regimes with scanning gate microscopy were investigated in detail in Refs.~\cite{Zhukov2011P, Zhukov2013} and \cite{Zhukov2012}, correspondently.

For a more comprehensive understanding of the influence of the charged AFM tip on the wire it is useful to discuss results of calculations made in the orthodox Coulomb blockade model \cite{Wiel2003} following Ref.~\cite{Zhukov2013}. In this model the InAs wire is considered to be divided into two sequentially connected quantum dots. The next main parameters are used to characterize this system: dot~1(2) to back-gate capacitance ($C_{1(2)BG}$), the mutual capacitance of the dots ($C_{m}$), and the capacitance of dot~1(2) to the left(right) contact ($C_{L(R)}$), the width of the energy levels of dot 1(2) ($\Gamma_{1(2)}$). The explicit expression of dot~1(2) to tip capacitance $C_{1(2)t}$ on the tip to dot distance can be found in \cite{Zhukov2013}. The presence of the charged AFM tip results in an additional charge induced in dot 1(2) $\Delta q \propto V_t C_{1(2)t}$. The expression $C_{1(2)}=C_{L(R)}+C_{1(2)BG}+C_{1(2)t}+C_m$ gives the the sum of all capacitances attached to dot 1(2). In our SGM experiments the widths of the energy levels of the quantum dots are considerably larger than the temperature smearing ($k_BT$, here $k_B$ is Boltzmann constant), so a Lorentz shape broadening of the energy levels can be used \cite{Nazarov1993}. We therefore use the formula $I \propto (\Gamma_1 ^2/(\Gamma_1^2+\mu_1^2))(\Gamma_2 ^2/(\Gamma_2 ^2+\mu_2^2))$ to calculate the current through the system, where $\mu_1$ and $\mu_2$ are the electrochemical potentials of dot 1 and 2, respectively \cite{Wiel2003, Zhukov2013}, the Fermi level energy of metallic contacts is assumed to be zero.

The result of the calculations for zero interdot capacitance ($C_m=0$) and equal level width of both dots  $\Gamma_1=\Gamma_2=1$~meV is presented in Fig.~6 in \cite{Zhukov2013}. The interpretation of the simulation result is rather simple. Two sets of concentrical circles of increased conductance through the wire comes from two dots presented in the model with their centers located in the centers of the corresponding dots. In the linear transport regime each circle denotes the position of the tip when the energy level of the corresponding dot coincides with the source and drain energy levels, i.e. the condition for the single electron tunneling conductance peak is satisfied. In the spots where different sets of circles crossing each other the energy levels of both dots are aligned with the source/drain energy levels and the conductance is maximized. It is easy to note the mirror-like symmetry of resulting model SGM scans calculations, the mirror is placed on the wire axis. 

In \cite{Zhukov2013} the observed wobbling of the Coulomb blockade conductance lines may be explained within the above model if the level width of dots are not equal, see Fig.~5 in \cite{Zhukov2013}. The interdot capacitance in this calculation was assumed to be zero ($C_m=0$). Additionally, the wobbling of the Coulomb blockade conductance lines may be observed if the interdot capacitance is comparable to the dot self-capacitance ($C_{L(R)}+C_{1(2)BG}$) and it is larger than $k_BT$. This type of wobbling is well known and it is observed in double dot systems based on two-dimensional electron gas such as in InAs-based heterojunctions \cite{Huefner2011}.

It is shown in \cite{Zhukov2011P} that in real samples the above mentioned mirror-like symmetry of the conductance lines may not been observed. Moreover, the symmetry changes dramatically with increasing back-gate voltage, see Fig.~4 in \cite{Zhukov2011P}. This behavior was attributed to a redistribution of the electrons density across the quantum wire. Thus, because of the electron density redistribution the tunneling rate through the blocking barriers might depend on the crossed electric field as well. This results in an altering of the total conductance of the wire \cite{Zhukov2011P}.

The experimental data obtained on sample with open contacts and a single tunneling barrier in the wire is presented in Fig.~2. The AFM tip alters the opacity of this barrier in the range of $250-300$\,nm, and within this range the distribution of the current in the wire is mostly recovered in the regime of linear electronic transport. Thus the elastic mean free path in the dominating diffusive channel is expected to be less than $l_e<300$~nm.
\begin{figure*}
\includegraphics [width=1.1\columnwidth]{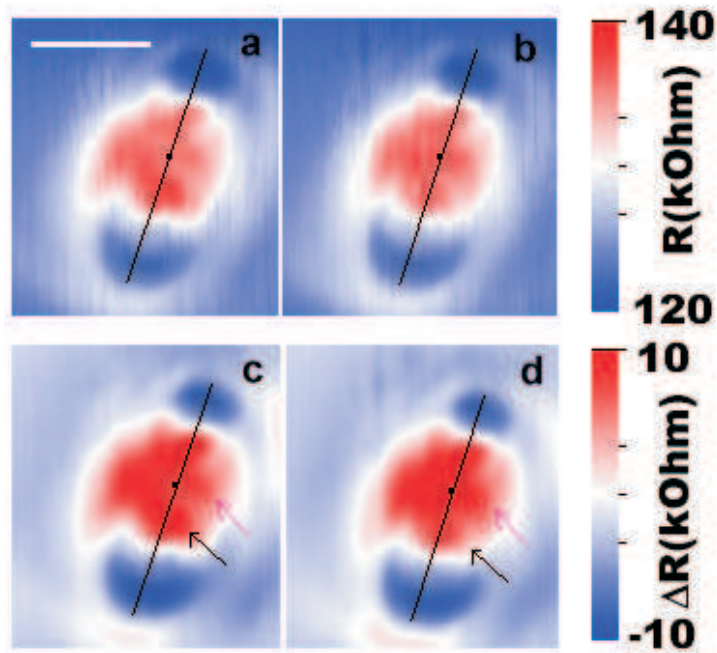}
\caption{(Color online) a) and b) SGM scans performed at $T=4.2$~K with a constant tip voltage of $V_t=$0~V and a back-gate voltages of $V_{BG}=13.96$~V and 13.98~V, respectively. c) and d) are obtained from a) and b) after smooth background subtraction. The black solid lines denote the wire and the dot correspond to the position of the tunneling junction. The horizontal white scale
bar in Fig.~a) corresponds to 1~$\mu$m, this scale is the same for all figures. Red and black arrows indicate the switching in the cross-like pattern placed over tunneling junction.} \label{Fig2}
\end{figure*}

In the InAs wire divided on two consequently connected closed quantum dots it is possible to observe the effect of negative differential conductance based on resonant tunneling through two consequently connected quantum dots \cite{Zhukov2012} similar to results obtained in lithography arranged samples based on a 2D electron gas \cite{Vaart1995}. With a charged AFM tip it is possible to adjust the size of this effect by fine tunning of the mutual energy levels of these quantum dots, see Fig.~3 in \cite{Zhukov2012}. Additionally, by employing the SGM technique it is possible to trace the moment of opening an additional conductive channel via higher energy level in the connected to drain quantum dot, see Fig.~4 in \cite{Zhukov2012}. 

Additionally, the analysis of the mismatch of the dot sizes ratio extracted from $V_{sd}$ vs. $V_{BG}$ conductance mapping and directly from SGM images made in \cite{Zhukov2011P} allows to evaluate the widths of the depletion regions of around 150~nm adjoined to the tunneling barrier from the both sides. As it was mentioned previously the increasing of the opacity of the tunneling barrier with increasing of the back gate voltage comes partially from the alteration of the depletion regions widths.

\subsection{3.2. Quasi-periodic oscillations}

If InAs wire has no tunneling barriers on the contact interfaces and no Coulomb blockade peaks in $1/R_{wire}$ vs. $V_{BG}$ are observed, see Fig.~4 in \cite{Zhukov2014SWComp}, than the main dependence of $1/R_{wire}(V_{BG})$ is linear. Aperiodic deviations of this main dependence with character size of $\sim 0.1(e^2/h)$ are usually attribute to universal conductance fluctuations (UCF) \cite{PLee,refBloe}. Here, $e$ is the elementary charge and $h$ is the Planck constant. From the temperature dependence of the UCF amplitude it is possible to extract the temperature dependence of the phase coherence length $\tau_\varphi(T)$ \cite{Hernandez,refBloe}.

Using a capacitor model it is possible to calculate the total amount of electrons $N \sim 2300$ added to the wire by changing the back-gate voltage from 0 to 10~V \cite{refWirt}. Knowing the geometrical sizes of the wire it is possible to calculate the mean concentration of the electrons  $n_{3D}\sim 1.1\times 10^{17}$~cm$^{-3}$. The reciprocal radius of Fermi sphere calculated by considering the InAs nanowire as a 3D structure is $2\pi/k_{F,3D}=2\pi/(3\pi^2 n_{3D})^{1/3} \sim 40$~nm, the elastic mean free path in the wire is $l_e=\sqrt[3]{3/8\pi n^2_{3D}}(h/e^2) (1/R_{wire})(L_{wire}/\pi r^2) \sim 50$~nm \cite{Roulleau,Dhara}. Thus, the statement: ``the InAs wire is a 3D structure in the diffusive regime, close to the Ioffe-Regel limit regime $k_{F,3D} l_e \sim 1$'' looks rather reasonable \cite{Zhukov2012SW}. For these nominally undoped InAs nanowires the main source of scattering is surface scattering, due to irregularities caused by stacking faults or due to surface contaminations \cite{refWirt}. But the evaluated value of elastic mean free path in InAs wire as 40~nm measured at $T=4.2$~K is in disagreement with data shown in Ref. \cite{Zhou}, where the measurements were performed at room temperature and the obtained value was more than 200~nm. In this experiment the potential distribution along the wire from source to drain was measured using the conductive tip of the AFM. 

At first glance no special structures might be observed in SGM scans if the tip to SiO$_2$ surface distance is more than 300~nm. If the tip-to-SiO$_2$ surface distance is of $\sim$200~nm the SGM scan might look like a random set of ripples with an amplitude of $0.1(e^2/h)$ i.e. similar to the size of universal contact fluctuations. Generally speaking, for confirmation of the ergodic theorem \cite{Anosov} it is more strict to use the value of fluctuations measured with SGM along the equipotential line than the value measured by sweeping the back gate voltage and compare this value to one measured by sweeping the magnetic field because the total number of electrons in the wire is constant in case of the SGM technique and only the distribution of scatterers is changed.

Thus, the observation of quasi-periodic oscillations with wavelength more than 250~nm in InAs nanowire using SGM was a kind of surprise \cite{Zhukov2012SW}. Nontrivial dependence of oscillation wave length ($l_{node}$) on the back-gate voltage allows to attribute them to electrons in the top subband, see Fig.~2 in \cite{Zhukov2012SW}. These electrons have small $k$ ($2\pi/k=l_{node} \sim l_{wire}$) and with such a small wave vectors they do not scatter on the wire wall resulting in a ballistic behavior of these electrons \cite{Zhukov2012SW}. Long range (more than 1$\mu$m) fluctuations of the potential are screened by the closely placed back-gate. The abrupt increasing of the oscillation period from ~250~nm to $\sim 1\mu$m is interpreted as taking the top subband to the ``disordered sea'' when $2\pi/k$ becomes comparable to wire diameter ($2\pi/k = l_{cut}$, see Fig.~3) with subsequent formation of the new top subband see Fig.~2d) and c) in \cite{Zhukov2012SW}. Rising $V_{BG}$ increases $k$ of these electrons, see Fig.~2 c) to a) in \cite{Zhukov2012SW}.  In the models of Wigner crystallization  of Luttinger liquid and Friedel oscillations a similar dependence of $1/l_{node}$ vs. $V_{BG}$ is expected \cite{Ziani2012}.
\begin{figure}
\includegraphics [width=0.9\columnwidth]{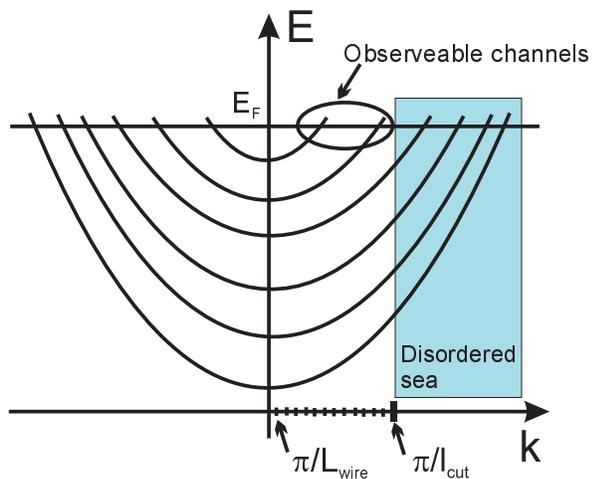}
\caption{
Schematic band diagram of the InAs nanowire, here $k$ is directed along the wire axis. The top subbands with small $k_F<\pi/l_{cut}$ form standing waves visualized in the SGM images. The lower lying subbands with $k_F>\pi/l_{cut}$
form the \emph{disordered sea}, and no individual channels can be distinguish
in this case. From Ref. \cite{Zhukov2012SW}.
}
\label{Fig3}
\end{figure}

The kinetic energy $E_k(N_{node}=3)=[(3/2h/l_{wire})]^2/(2m^*)=21\,\mu$eV is basically less than $k_BT$ even for the three-nodes resonant state. Here, $m^*=0.023\,m_e$ is the effective electron mass in InAs and $m_e$ is the free electron mass. The thermal length $L_T(N_{node}=3)=h^2/(2\pi m^*k_BT(2l_{wire}/3))\sim 30$\,nm is considerably smaller than the length of the wire $l_{wire}=2.6$\,$\mu$m. It is worth noting, the energy difference between the $N_{node}=2$ and $N_{node}=3$ resonance states is slightly less than $k_BT$ because both states are resolved simultaneously at a back-gate voltage $V_{BG}=9.62$~V (cf. Fig.~1f) in \cite{Zhukov2014SWComp}). Additionally, the Coulomb energy of the electron-electron interaction, $V_C(N_{node}=3)=2e^2d_d^2/(4\pi\epsilon_0\epsilon)(l_{wire}/3)$, even for the three-nodes resonant state is less than $k_BT$ as well. Here, $\epsilon=15.15$ is the static dielectric constant in InAs, $d_d \sim 150$~nm is the distance from the center of the wire to the doped back-gate, and $\epsilon_0$ is the vacuum permittivity. Thus, the ability of Wigner crystallization, Friedel oscillations and standing wave scenarios looks rather doubtful for $N_{node}=2$ and 3. 

However, the Coulomb interaction between the electrons becomes comparable to the temperature ($V_C \sim 0.16$\,meV) only at $V_{BG}=10.40$\,V when the average distance between the electrons is around 250\,nm. We may speculate that the
formation of a Wigner crystal in the top subband happens at this back-gate voltage with a $2k_F$ to $4k_F$ transformation of the oscillations.

As it was found previously \cite{Matveev2007}, a Wigner crystal may posses a so called ``zig-zag instability'' when the electron concentration increases. A similar behavior was predicted and observed for electrons on a He$^4$ surface \cite{Chaplik1980}. For our SGM experimental data an instability of this kind might result in a special slalom-like path for the conductive electrons. Such kind of paths are shown in Fig.~4a) measured in InAs wire without defects and Fig.~4b) measure in InAs wire with one tunneling barrier. observation of such paths is an additional argument toward statement of Wigner crystal formation if $l_{node} \sim 250-300$~nm observed in both SGM scans, see Fig.~4. Note, the tunneling barrier in InAs wire does not destroy the formed structure, see Fig.~4b). 
\begin{figure}
\includegraphics [width=0.95\columnwidth]{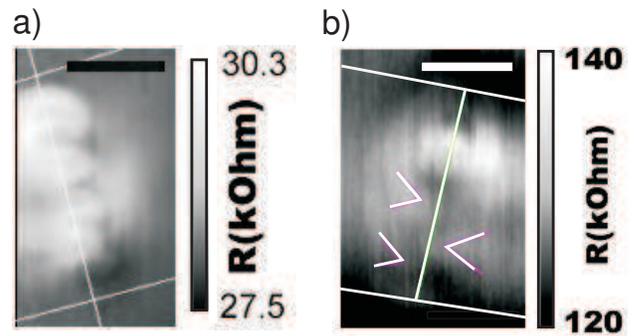}
\caption{
a) is the SGM image of an InAs wire (see Fig.~1a)), $V_t=0$~V, $V_{BG}=8.80$~V. b) is the SGM image of an InAs wire (see Fig.~2)), $V_t=0$~V, $V_{BG}=13.50$~V. The white solid lines denote the wire axis and the boundaries of source and drain electrodes. The horizontal scale bars corresponds to 1~$\mu$m. Angles in b) indicate knees of the slalom-like curve. 
}
\label{Fig4}
\end{figure}

These slalom-like paths are rather unstable and they crash easily with changing back-gate voltage for 50~mV and, additionally, they are quite hard to observe. The reason for it is clear, the distance of 250-300~nm in between electrons for $V_C \sim k_BT$ is close to $l_{cut}$ and also this length is comparable to the resolution of our experimental setup of $\sim 150-200$~nm.

Taking into account the scale of kinetic energy for $N_{node}=2$ mentioned previously, the investigation of the stability of the observed structure looks rather important \cite{Zhukov2014SWComp}. The resulting experimental data is presented in Fig.~2 in \cite{Zhukov2014SWComp}. No significant deviations of the node positions and the amplitude of the oscillations are found as long as $eV_{SD}\le k_BT$. But the application of a large current of $I_{AC}=50$~nA resulting in $eV_{SD}\sim 1.6\,$meV$\,\gg k_BT$ suppresses the oscillations and decreases the resistance of the whole wire (cf. Fig.~2e) in \cite{Zhukov2014SWComp}). 

In paper \cite{Zhukov2014SWComp} the conductivity of the electrons in the ``diffusion sea'' was established as well. As it was mentioned previously, the charged AFM tip influences the electron locally in case of a flat potential background if $2\pi/k$ is comparable to the  electron-to-tip distance \cite{Zhukov2009}. Thus the experimental evidence of influence of the electron in the ``disordered sea'' for a 200-220~nm tip-to-surface distance experimental setup is expected. The resulting set of SGM scans is presented in Fig.~3 in \cite{Zhukov2014SWComp}. 

The pattern of the ripples observed in Fig.~3d) to f) in  \cite{Zhukov2014SWComp} changes essentially when $V_{BG}$ rises just for 20\,mV while SGM images are obtained with scanning tip placed close to the nanowire ($l_{tip} \sim d$), here $d$ is the wire diameter.  In contrast to that, the width of the step as a function of $l_{node}(V_{BG})$ is more than hundred millivolts. Thus, Fig.~3 in \cite{Zhukov2014SWComp} presents a solid confirmation that electrons in the disordered sea are not localized.

The irregular pattern observed in the SGM scans governed by the ``disordered sea'' electrons has the smallest typical length scale of 200~nm. This is in accordance with the expected spatial resolution of the experimental setup for $h_{tip}=220$~mn, and the amplitude of the deviations of the resistance (cf. Fig. 3g) in \cite{Zhukov2014SWComp}) is comparable with the amplitude of the universal conductance fluctuations shown in Fig.~4a) (inset) in \cite{Zhukov2014SWComp}.

Generally speaking, it is possible to trace the formation of new top subband in the wire performing SGM scans as it has been done in \cite{Zhukov2012SW}. But this procedure is extremely time consuming. In Ref.~\cite{Zhukov2014SWComp} the new method of detection of formation of top subband was suggested.

Let us consider the mechanism of top subband formation in details following Ref.~\cite{Zhukov2014SWComp}. By increasing the back-gate voltage just up to the instance of the new subband formation the electrons are loaded simultaneously to the disordered sea and to the subband. The electrons loaded to the new subband are blocked because of the potential barriers at the wire to contact interfaces forming a semi-opened quantum dot. This dot decreases the total conductance of the wire at certain range of back-gate voltage, namely from $V_{BG}$ of quantum dot formation to the value of the back-gate voltage when the barriers become transparent. Let us call the center of this range $V_{BG1}$. It is possible to slightly alter $V_{BG1}$ with the charged AFM tip. Taking into account that the electrons in the quantum dot are concentrated in the center of the wire, $V_{BG1}$ as a function of the tip position along the wire must have one maximum \cite{Zhukov2009,Boyd2011}. This maximum is traced by the dashed line in Fig.~4b) in  \cite{Zhukov2014SWComp}. Thus, three subbands are formed marked as I, II and III in the region of back-gate voltage of
0\,V$\leq V_{BG}\leq 12$\,V at $V_{BG} \approx 2.7$, 7.5 and 10.5\,V, respectively. There is some discrepancy for back-gate voltages less than 1~V in the determination of the formation of the second subband comparing to the data shown in Ref. \cite{Zhukov2012SW}, where the formation was determined directly
from SGM scans. It originates from the hysteresis in $R_{wire}(V_{BG})$ of the sample.

Observing the formation of three subbands in the 0~V$<V_{BG}<12$~V back-gate range means that the total number of loaded subbands may be evaluated as 4 or 5 and, thus, the total number of conductive electrons loaded into the wire containing 1D electrons of the top subband and the 3D electrons of the "disordered sea" is less than 100. The total number of electrons added to the wire calculated from the capacitance is  around 2500 \cite{Zhukov2012SW}. This means that most of the electrons in the wire are trapped probably because of interface states charged by changing the back-gate voltage. Thus, all electrons in the InAs wire are of three types, namely, 1D conductive electrons in the top subband, 3D conductive electrons in the "disordered sea" and localized electrons in the traps. This also means that the value of the mean free path must be recalculated as well. It seems $l_e$ is actually larger than the typically used value of 40~nm for the disordered sea, but might be smaller than 300~nm according to experimental results presented in Fig.~2.

\section{4. Conclusion.}

In conclusion, we performed a set of measurements using SGM technique on InAs nanowires with and without internal tunneling barriers (weak junctions). The main results are:

-- The ratios of QD sizes extracted from conduction maps as function of $V_{SD}$ and $V_{BG}$ and from SGM measurements are in good agreement. To explain the deviation of the conductance while the charged tip moved along the equipotential lines the redistribution of the electrons density across wire in weak junction must be taken into account. Within the range of $\sim 300$~nm the current distribution in the wire is mostly recovered in the regime of linear electronic transport. Thus the elastic mean free path in the disordered sea is expected to be less than $l_e<300$~nm.

-- If the wire is split into two quantum dots connected in series a wobbling and splitting of the Coulomb blockade conductance peak lines is observed. Both effects are explained as the interplay of the conductance of two quantum dots being present in the system. Simulations of the wobbling in the frame of the orthodox theory of Coulomb blockade involved two quantum dots connected in series demonstrate good agreement with experimentally obtained SGM data.

-- Adjusting the opacity of the tunneling barrier between the two closed quantum dots by means of a charged AFM tip, a negative differential conductance is induced in the system. The shift of the $V_{SD}$ regions where a negative differential conductance occurs while a different tip voltage induces an alteration of the mutual energy level positions of the dots. It is shown that a suppression of the negative differential conductance can be attributed to an additional conductivity channel through the energy state of the neighbor quantum dot if the tip moves along a particular equipotential trajectory.

-- Quasi-periodic oscillations are observed in InAs nanowire ($R_{wire} \sim 30\,$k$\Omega$). The period and the amplitude of them are investigated in detail as a function of back-gate voltage in the linear and non-linear regime. None of the scenario such as Friedel oscillations, Wigner crystallization or standing wave looks applicable to explain the origin of observed oscillations at $N_{node}=2$ and 3. Observed slalom-like path was attributed to zig-zag instability of provisional Wigner crystal.

-- We demonstrate the influence of the tip-to-sample distance on the ability to locally affect the top subband electrons as well as the electrons in the disordered sea. We suggest a new method to evaluate the number of conductive electrons in an InAs wire. This method results in the conclusion that most of the electrons added to nanowire conductive band on applying a positive back-gate voltage are trapped.

\section{Acknowledgments}

This work is supported by the Russian Foundation for Basic
Research, programs of the Russian Academy of Science, the Program
for Support of Leading Scientific Schools, and by the
International Bureau of the German Federal Ministry of Education
and Research within the project RUS 09/052.

{}


\end{document}